\documentclass[final,3p,times,twocolumn]{elsarticle}

\usepackage{amssymb}

\usepackage{url}
\usepackage{graphicx}
\usepackage{subfig}
\usepackage{amsmath}

\journal{Nuclear Instruments and Methods A}

\begin{document}

\begin{frontmatter}



\title{Simulation of the response of a diamond-based radiation detector to ultra-short and intense high-energy electron pulses}


\author[INFN]{Y.~Jin}
\author[INFN]{P.~Cristaudo}
\author[INFN,UniTS]{A.~Gabrielli}

\address[INFN]{INFN, Sezione di Trieste, I-34127 Trieste, Italy}
\address[UniTS]{Dipartimento di Fisica, Universit\`a di Trieste, I-34127 Trieste, Italy}

\begin{abstract}
Single-crystal synthetic diamond sensors have been widely used in radiation dosimetry and beam diagnostics. The foreseen harsh radiation environment in electron-positron colliders at the luminosity frontier requires a thorough investigation of diamond's response to large radiation burst, in particular, to intense high-energy electron pulses. In this article, a two-step numerical simulation approach (Sentaurus~+~LTspice) is proposed to explore this topic. Time response of the diamond detector is simulated via TCAD-Sentaurus while the transmission effect of the electronic circuit is taken into account using LTspice. Good agreement is observed between results of the numerical simulation and preliminary experimental data from detector's exposure to high-energy sub-picosecond electron pulses, on both the amplitude and the shape of the induced signals. This simulation combination is a novel approach to designing and optimising diamond detectors for radiation and beam loss monitoring in particle physics experiments.
\end{abstract}


\end{frontmatter}



\section{Introduction}
\label{sec:introduction}

Single-crystal diamond sensors synthesized by chemical vapor deposition (sCVD) have been widely used in radiation dosimetry and beam diagnostics~\cite{ref_natural,CANALI1979,ref_cms,ref_atlas,ref_babar}. We have developed and installed a diamond-based radiation monitor and beam abort system~\cite{ref_performance} in the Belle II experiment~\cite{ref_BelleII} at the SuperKEKB asymmetric-energy electron-positron collider~\cite{ref_SKB}. With the push towards higher instantaneous luminosity of the collider, beam backgrounds become more severe, and high radiation bursts induced by beam losses can cause localized damage on essential Belle II sub-detectors and SuperKEKB components. Thus, the harsh radiation environment at the luminosity frontier urges a thorough investigation of diamond's response to large and fast radiation transients such as those induced by intense high-energy electron pulses.

In its simplest configuration, a typical diamond detector is obtained by the deposition of two metallic electrodes on opposite faces of a flat crystal. When a voltage bias is applied by an external circuit, the electrons and holes, liberated by ionization, drift towards the electrodes, inducing a current signal in the external circuit. Thanks to their high charge carrier mobility, diamond detectors can generate very fast signals. However, in the case of large ionization energy deposited in diamond, the large concentration of excess charge carriers in the diamond bulk gives rise to plasma effects that affect the rising time and amplitude of the signal. In addition, further signal reflection and distortion occur due to the external circuit. The determination of the signal is an interplay between charge carrier transport in diamond bulk and signal transmission in the circuit. As a result, unlike the steady continuous radiation that has been well calibrated for our diamond system~\cite{ref_cali}, the dose rate estimated on radiation bursts that have spikes-like time dependence will encounter non-linear effects and lead to an underestimated dose value. 

Successful attempts have been made using Technology Computer-Aided Design (TCAD) software~\cite{ref_synopsys} to investigate some properties of diamond devices, e.g., the charge collection efficiency under different bias voltages~\cite{ref_perugia}, transient current caused by single $\alpha$-particle hit~\cite{ref_kek}. In light of these successful applications, we carry out the simulation of the detector embedded in a simplified external circuit via ``mixed-mode" TCAD-Sentaurus with its SPICE implementation. Subsequently, to take into account the circuit effects with a detailed model overcoming the limitations of the ``mixed-mode" TCAD, the results of the first step are input to LTspice simulator~\cite{ref_ltspice}. This new factorized numerical simulation approach is proposed to explore the time response of a diamond device to ultra-short and intense high-energy electron pulses. Such a two-step simulation (Sentaurus~+~LTspice) can interpret the measured data and extrapolate the performance of diamond device in similar radiation environments. 

An experimental program is ongoing at the electron-linac of the FERMI free-electron laser in Trieste~\cite{ref_FEL}, to study the response of the diamond detectors to the irradiation by intense, sub-picosecond bunches of electrons accelerated to about 1~GeV. We compare our simulations with preliminary data from Ref.~\cite{ref_ICHEP}.

The experimental setup is briefed in Section~\ref{sec:experiment}. The two-step simulation approach is elucidated in Section~\ref{sec:workflow}. The two steps with regard to the response of the diamond detector and the effect of the electronic circuit are demonstrated in Section~\ref{sec:single} and Section~\ref{sec:circuitry}, respectively. A validation of the approach has been devised in Section~\ref{sec:alpha}. Final results of the numerical simulation overlaid with the experimental data are shown in Section~\ref{sec:results}.

\section{Experimental setup}
\label{sec:experiment}

We briefly describe here the diamond detectors developed for the Belle II experiment and the beam test facility set up to characterize their behaviour under extreme irradiation conditions.

\subsection{Diamond detectors}
\label{subsec:diamond}

The diamond sensors consist of $(4.5 \times 4.5 \times 0.5)$~mm$^3$ high-purity sCVD diamond crystals provided by Element Six Ltd~\cite{ref_e6} and two $(4.0 \times 4.0)$~mm$^{2}$ electrodes on opposite faces. The electrodes are made of Ti/Pt/Au layers with $(100$+$120$+$250)$~nm thickness, processed by CIVIDEC~\cite{ref_cividec}. More details can be found elsewhere~\cite{ref_performance}. During the measurements, one electrode of the diamond sensor is connected to a custom-made HV power supply~\cite{Bepo}, the other is connected to a LeCroy HDO9000 oscilloscope~\cite{ref_scope}.

\subsection{Beam facility}
\label{subsec:fermi}

Collimated electron bunches of about 1~ps duration with energy up to 1.5~GeV are available at the linac of the FERMI free-electron laser in Trieste~\cite{ref_FEL}. Bunch charge can be tuned from tens to 1000~pC and transverse size down to about $0.1$~mm. The first
data taking is carried out using 0.9~GeV electron bunches of 35~pC charge ($<$1$\%$ of bunch charge at SuperKEKB) and 120~\textmu m transverse size.

\section{Simulation workflow}
\label{sec:workflow}

TCAD-Sentaurus is regarded as the industry standard for the numerical simulation of the electrical characteristics of silicon-based and compound semiconductor devices~\cite{ref_synopsys}. The predictive power of its numerical calculation on the response of semiconductor devices to external electrical, thermal, and optical sources has been verified via numerous applications in the semiconductor industry. Using TCAD-Sentaurus, a series of sequential processes of the diamond detector are simulated, including a radiation beam interacting with the diamond crystal, the creation of electron-hole pairs, the drift of charge carriers, and the evolution of the induced voltage drop on electrodes. In addition, the space-time evolution of the concentration of charge carriers inside the diamond crystal is also obtained. These processes are regarded as the first step of the simulation workflow and are described in Section~\ref{sec:single}.

Though a SPICE-like utility is embedded in the TCAD-Sentaurus package, its functionalities and models are not adequate to fully take into account the effect of transmission cables, which plays a leading role in dealing with fast pulses. To overcome these limitations, LTspice, the most widely used SPICE software in the industry~\cite{ref_ltspice} is employed to take over as the second step. In our model described in Section~\ref{sec:circuitry}, the diamond detector is implemented by a combination of a voltage source in series with a resistor and in parallel with a capacitor. The simulated result of the evolution of voltage drop on electrodes from the first step serves as an input to the LTspice simulation for the voltage source. The resistance in the diamond detector model is obtained from the time evolution of voltage and current in TCAD. Coaxial cables, power supply, and oscilloscope all are properly modeled to take into account the transmission effects on the electrical signal such as reflection, attenuation, and distortion.

\section{Response of the diamond detector}
\label{sec:single}

The First step of the simulation is described here, including our choices for the physical parameters and models in TCAD and the generation of excess charge carrier by the ionizing electron beam.

\subsection{Diamond physical parameters}
\label{subsec:property}

The default values of parameters for the physical properties of diamond in the database of TCAD-Sentaurus are set according to Ref.~\cite{ref_web}. In addition, we update several sets of parameters. The mobility and saturation velocity of charge carriers are updated with values from our preceding measurements~\cite{ref_cali}. In the case of electrons (holes), $\mu_{0} = 1.72 \times 10^{3}$~cm$^{2}/$Vs ($2.05 \times 10^{3}$~cm$^{2}/$Vs), $v_{sat}=0.85 \times 10^{7}$~cm/s ($1.29 \times 10^{7}$~cm/s), where $\mu_{0}$ is the mobility extrapolated to low field and $v_{sat}$ is the saturation velocity at high field. Lifetimes of both charge carriers are reported as 2~\textmu s in the handbook~\cite{ref_handbook} from the producer Element Six Ltd. However, measurements of several groups using diamond crystals from Element Six Ltd give diverse results ranging from tens of nanoseconds to one microsecond~\cite{lifetime1, lifetime2, lifetime3, lifetime4}. This parameter will be revisited at Section~\ref{subsec:output}.

\subsection{Models in TCAD-Sentaurus}
\label{subsec:models}

\begin{figure}
	\begin{center}
		\includegraphics[width=6.5cm]{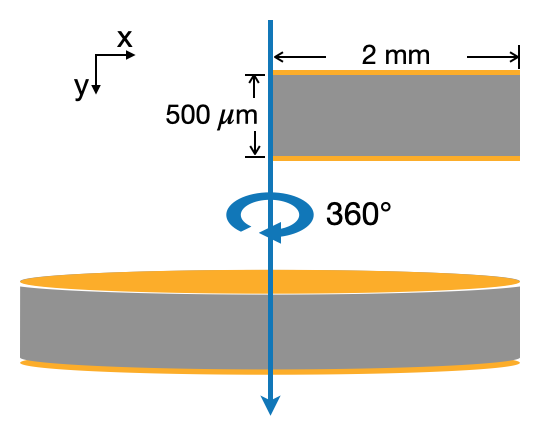}
		\caption{Geometry of the diamond detector in TCAD-Sentaurus. To make meshing easier, thickness of electrodes are set as 10~\textmu m, much thicker than their real dimension. Incident electron bunches traverse the detector along the arrow.}
		\label{fig:geometry}
	\end{center}
\end{figure}

To balance precision and computation time, the \textit{2D-Cylindrical Model} is adopted to describe the geometry of the diamond detector as shown in Figure~\ref{fig:geometry}. This model assumes that the behavior of the device is identical with respect to the azimuthal angle, so that a 3D device can be described in a 2D coordinate system.

Given the extremely low concentration of impurities (Nitrogen $<$ 5~ppb; Boron $<$ 1~ppb) in the synthetic sCVD diamond crystal with respect to that of the foreseen excess charge carriers generated by the intense electron bunch, we set the diamond crystal as non-doped. The \textit{Hydrodynamic} model is employed for carrier transport. Mobility of charge carriers is formulated as 
\begin{equation}
v_{d}(E) = \frac{\mu_{0}E}{1+ \frac{\mu_{0}E}{v_{sat}}}
\end{equation}
using option \textit{HighFieldSaturation}. The \textit{ElectricField} model is chosen for driving force. 
For recombination process, several models in TCAD-Sentaurus can be of use, e.g., \textit{Shockley-Read-Hall Model}, \textit{Radiative Recombination Model}, and \textit{Exciton Dissociation Model}.
However, at the time being, relevant experimental measurements are not sufficient to parameterize each process. As a trade-off, for a first trial, only \textit{Shockley-Read-Hall Model} is activated while most parameters in its formula are set to zero. Thus the model degenerates to a fixed value of lifetime regardless of concentration and temperature.
Note that due to the high density of charge carriers, dedicated models for the recombination at the surface can be neglected. The contact between diamond bulk and Ti/Pt/Al electrodes is set as ohmic~\cite{ohmic1992, ohmic2000}. 

In the application of radiation monitor, to reduce stochastic uncertainty, avalanche process is disfavored. On the other hand, a sufficient bias voltage is required to guarantee a full charge collection efficiency. In view of these considerations, our diamond system at Belle II is operating with 100~V bias voltage, which corresponds to an electric field intensity 0.2~V/\textmu m. For the study of transient response of the diamond device, we also extend to cases with 50~V and 150~V bias voltage applied on electrodes.

\subsection{Settings for electron beam}
\label{subsec:beam}
Using TCAD-Sentaurus' radiation model, \textit{Heavy Ion}, it is possible to parameterize the distribution of the energy deposited by the intense electron beam in the diamond crystal. The incident beam is normally traversing the diamond crystal along the axis, $x=0$, with a Gaussian transverse profile of $\sigma = 120$~\textmu m. The starting time of the injection is set to 1~ns. Due to the thinness of the detector, bremsstrahlung photons stemming from the radiative stopping power barely interact with diamond crystal (deposit energy $\sim$ 10~keV). Only the collision stopping power of diamond, $S(E)$, is considered to be responsible for the generation of charge carriers. According to a recent study~\cite{ref_deposit2021}, for 0.9~GeV electrons, assuming the density of diamond is 3.5~g$/$cm$^{-3}$, $S(E) = $ 7.0~MeV$\cdot$cm$^{-1}$. This value agrees with the database of NIST~\cite{ref_deposit}. The linear energy transfer (LET) is thus given by:

\begin{equation}
LET = \frac{S(E) \cdot N_{e} \cdot q_{e} }{E_{eh}} = 1870 ~\textrm{pC} / \textrm{$\mu$m},
\end{equation}
where $E_{eh} = 13.1~$eV is the energy to create an electron-hole pair in diamond, $N_{e}$ is the number of electrons in one bunch ($N_{e} = 2.18 \times 10^8$ for a 35~pC bunch), and $q_{e}$ is the elementary charge. In addition, the assumption of collision stopping power only has been cross-checked via a FLUKA simulation~\cite{ref_FLUKA, ref_FLUKA2}, which gives a mean deposit energy of 0.30~MeV in agreement with 7.0~MeV$\cdot$cm$^{-1}~\times$~0.05~cm.

\subsection{Outcome of the diamond detector}
\label{subsec:output}

\begin{figure}
	\begin{center}
		\includegraphics[width=5cm]{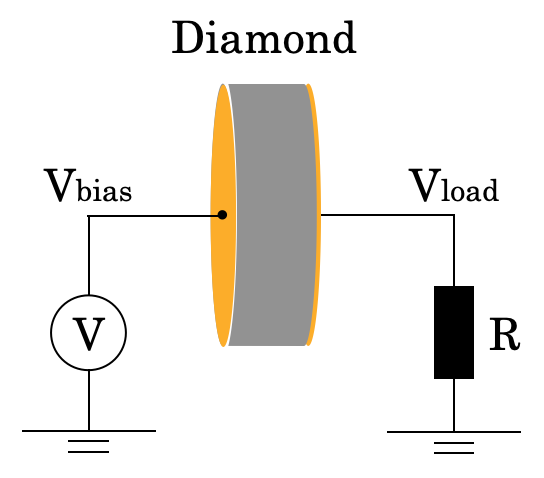}
		\caption{Sketch of the diamond detector in TCAD-Sentaurus. One electrode is biased by a voltage source, the other is connected to the ground via a load resistor (50~$\Omega$).}
		\label{fig:DD}
	\end{center}
\end{figure}

To apply the high bias voltage on electrodes, rather than setting a fixed value on electrodes, a simple circuit model is employed in the ``mixed-mode" simulation of TCAD-Sentaurus package, in which one electrode is connected with a DC voltage source (V$_{\text{bias}}$) and the other electrode is connected to the ground via a load resistor R, as shown in Figure~\ref{fig:DD}. The simulated evolution of the voltage on the latter electrode (V$_{\text{load}}$(t)) is input to LTspice.

\begin{figure}
	\begin{center}
		\includegraphics[width=7cm]{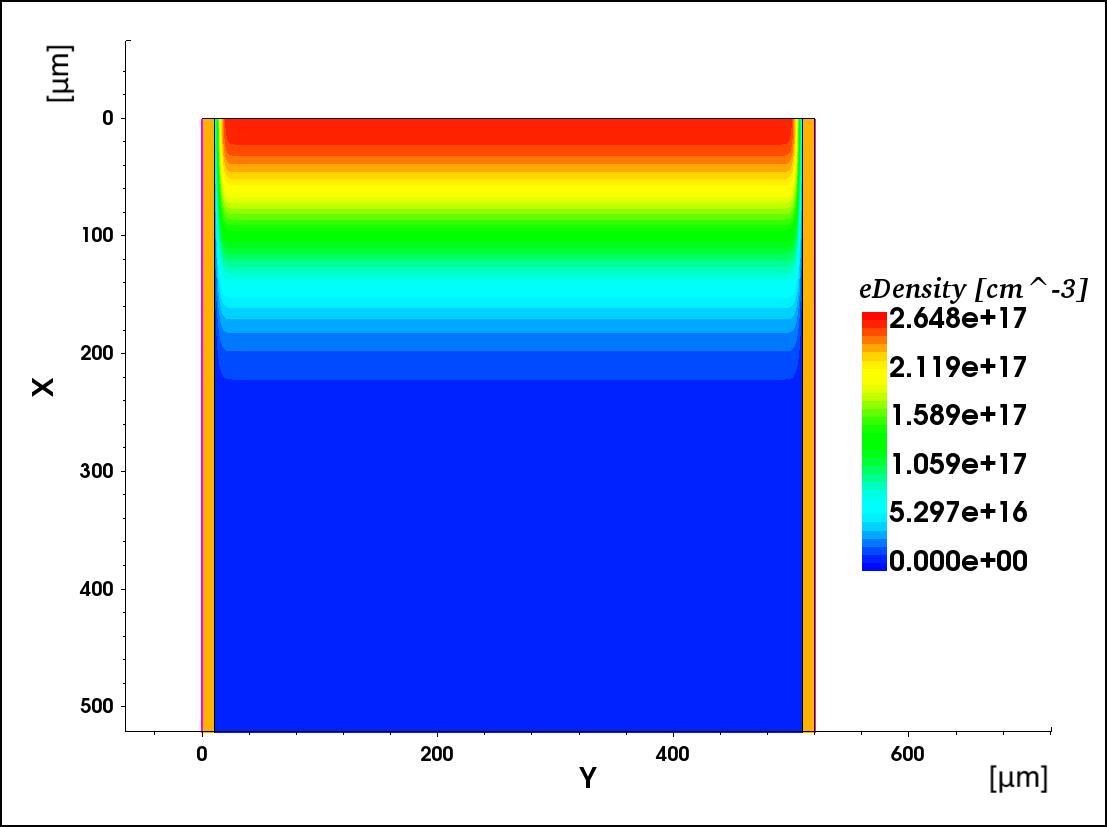}
		\caption{A zoom-in plot of the concentration of negative charge carriers (electrons) in the diamond bulk near the trajectory of incident electron bunches. The electron bunches are traversing along $X=0$. The dimension in x axis extends to 2~mm. The bias voltage in this case is 100~V. Similar distributions are obtained for the case of 50~V bias and 150~V bias.}
		\label{fig:eDen}
	\end{center}
\end{figure}

In addition, the evolution of the concentration of charge carriers in the diamond bulk is simulated by TCAD-Sentaurus as well. Figure.~\ref{fig:eDen} shows the concentration of negative charge carriers (electrons) in the center of diamond bulk at 1~ns after the traversing of beam. Similar distribution is obtained for positive charge carriers (holes). The concentration near the beam passage reaches up to $10^{17}~cm^{-3}$, at which excitonic recombination involving phonons take place~\cite{ref_recombine}. Also considering the temperature increase of diamond bulk due to the 50~Hz irradiation, which shortens the charge carrier lifetime, the value of lifetime in simulation is set to 50~ns, taking the lower bound of the aforementioned experimental results.

\begin{figure}
	\centering
	    \includegraphics[width=7.5cm]{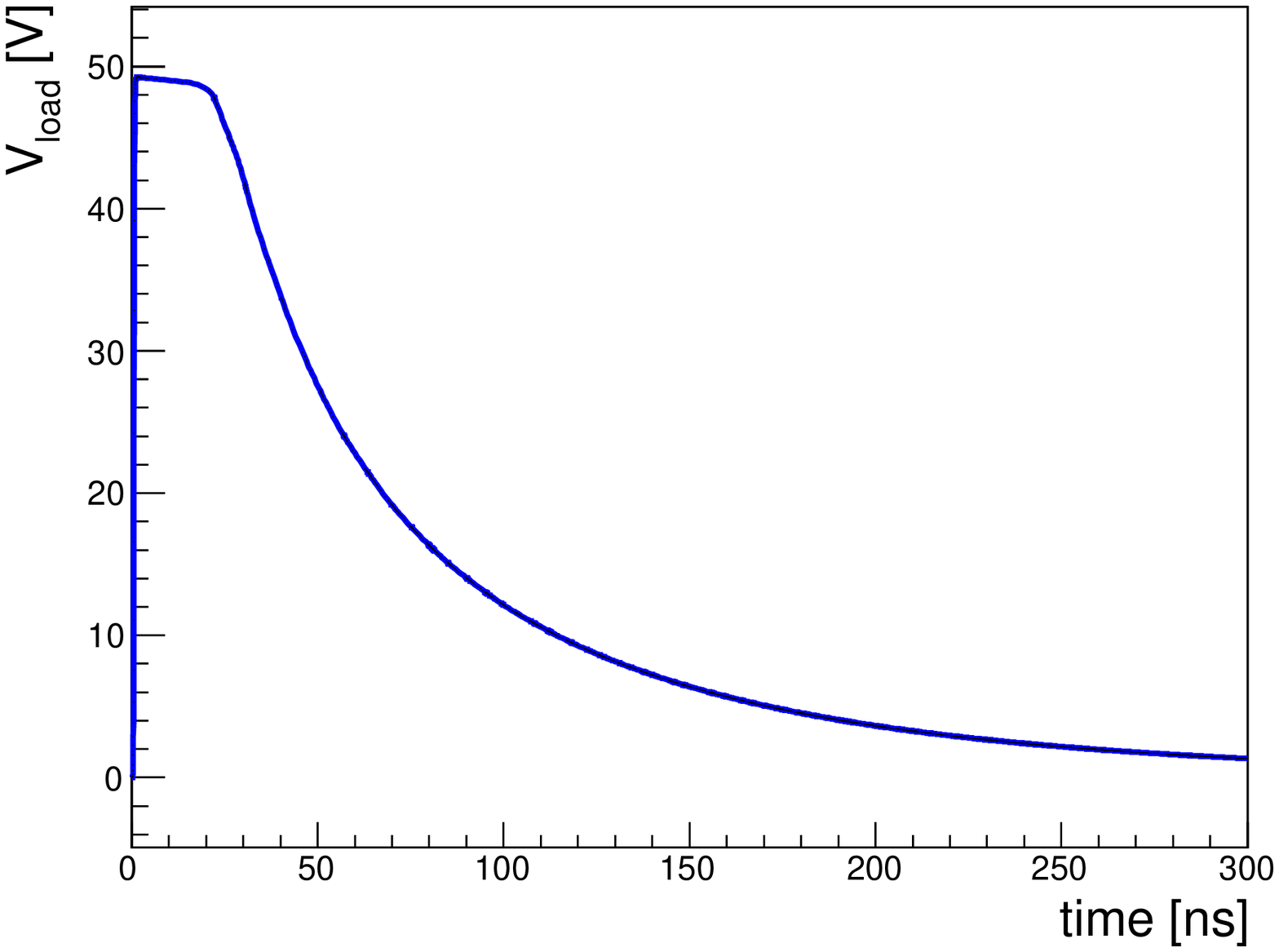}
	    \includegraphics[width=7.5cm]{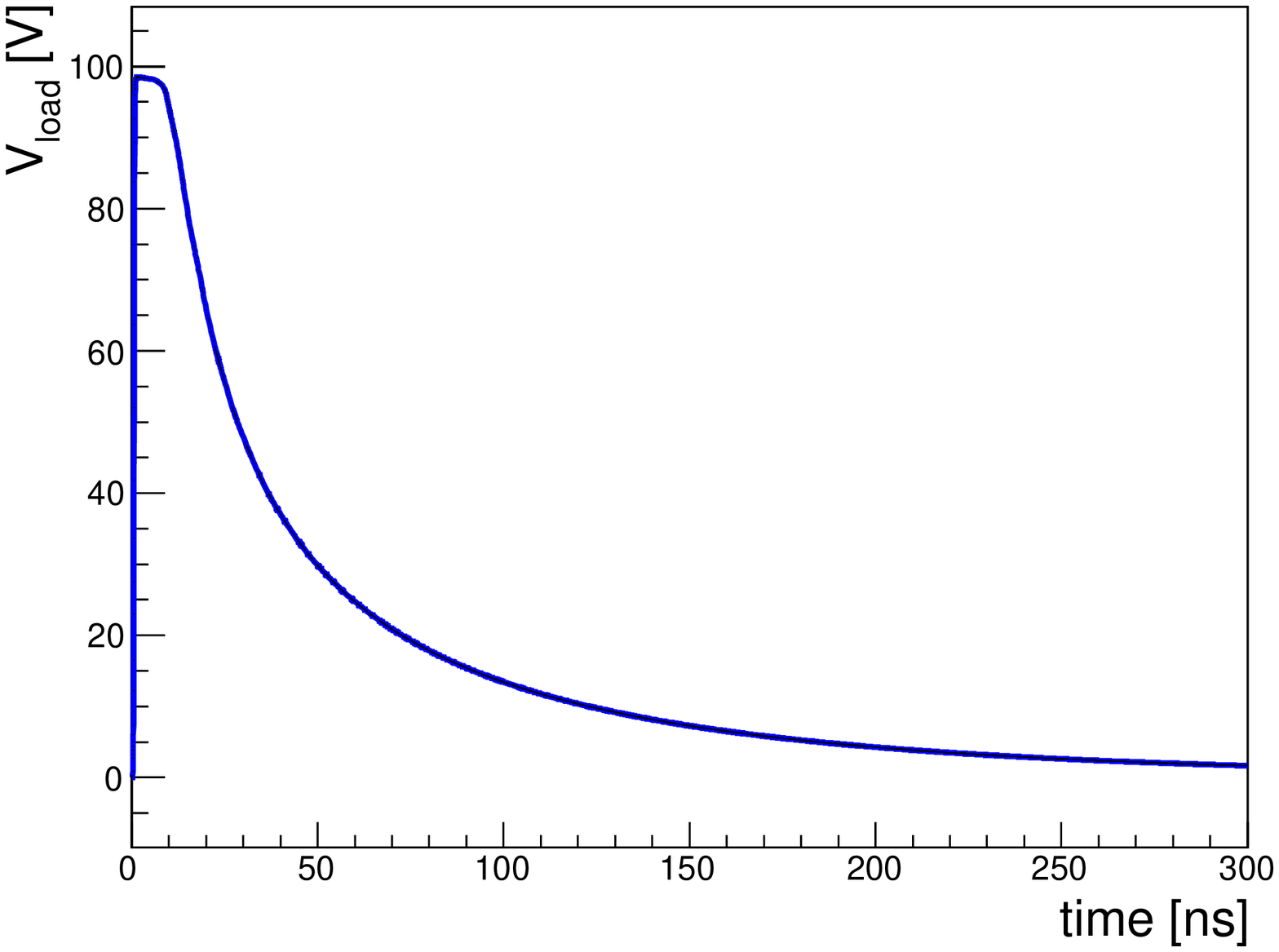}
	    \includegraphics[width=7.5cm]{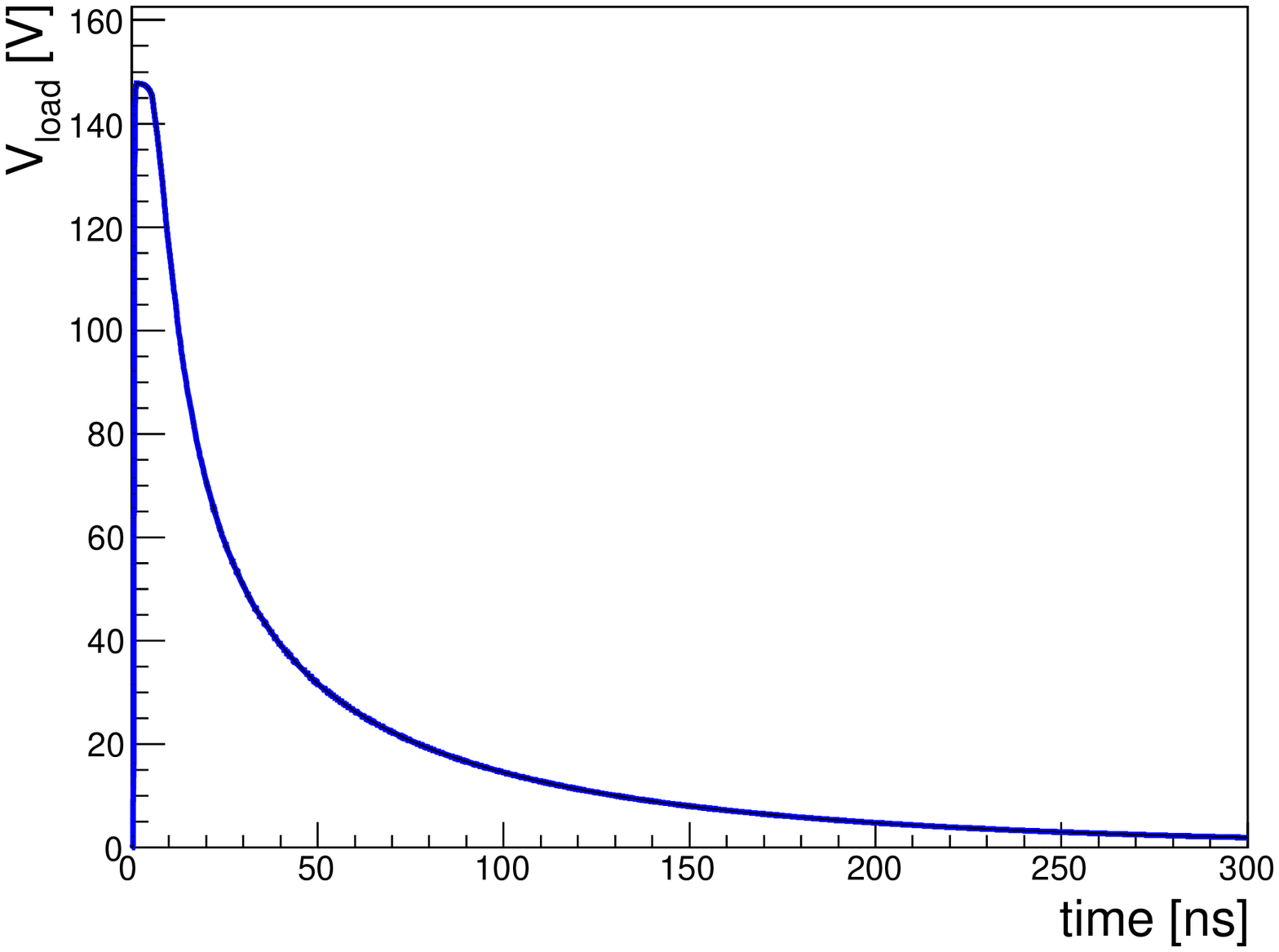}
	\caption{Voltage of the electrode connected to the ground as a function of time, V$_{\text{load}}$(t). Bias voltage is applied on the other electrode. Top: V$_{\text{bias}} =$ 50~V. Middle: V$_{\text{bias}} =$ 100~V. Bottom: V$_{\text{bias}} =$ 150~V.}
	\label{fig:outcome}
\end{figure}

\begin{figure}
	\begin{center}
		\includegraphics[width=7cm]{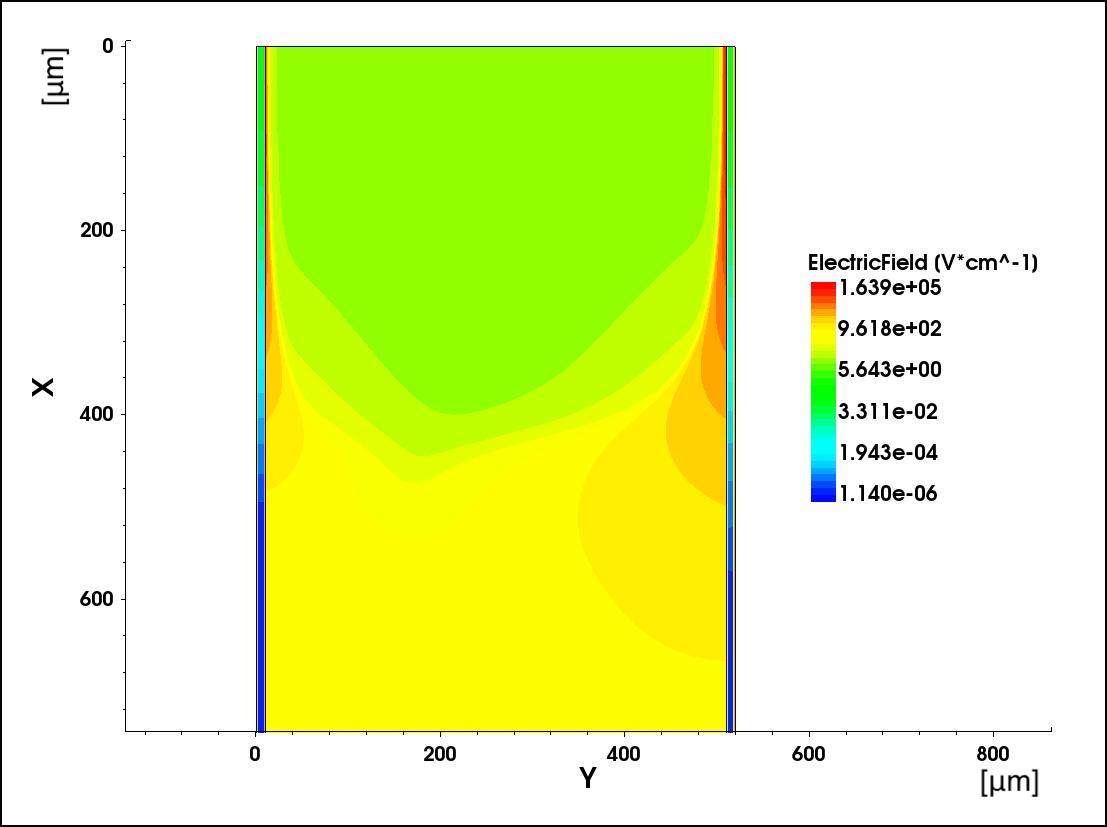}
		\caption{A zoom-in plot of the magnitude of electric field in the diamond bulk near the trajectory of incident electron bunches. Two types of charge carriers drift towards opposite directions leading to space charge layers near the electrodes with sufficient net charge to cancel off the external electric field. The green region is screened and has a relatively low magnitude of electric field. This is the simulated result of V$_{\text{bias}} =$ 100~V, at 20~ns after irradiation. Similar distributions are obtained for the case of 50~V bias and 150~V bias.}
		\label{fig:screening}
	\end{center}
\end{figure}

As demonstrated in Figure~\ref{fig:outcome}, it should not come as a surprise that a ``knee" feature appears in the voltage evolution. Immediately after the traversing of the electron beam, a huge amount of electron-hole pairs are liberated. Driven by the external electric field, charge carriers of two types separate and move towards opposite directions. Owing to the large concentration, these mobile charge carriers induce a transition of diamond into a ``conductor" and the voltage difference (V$_{\text{bias}} -$ V$_{\text{load}}$) across the two electrodes drop almost to zero. Subsequently, with the further separation of the two types of charge carriers, thin layers of only one type of charge carriers emerge near two electrodes, whose net charges are sufficient to establish an internal electric field that can cancel off the external field as sketched in Figure~\ref{fig:screening}. This screening phenomenon blocks the draft of charge carriers in the middle of diamond bulk so V$_{\text{load}}$ only decreases slowly.

\section{Effect of the electronic circuit}
\label{sec:circuitry}

\begin{figure*}
	\begin{center}
		\includegraphics[width=15cm]{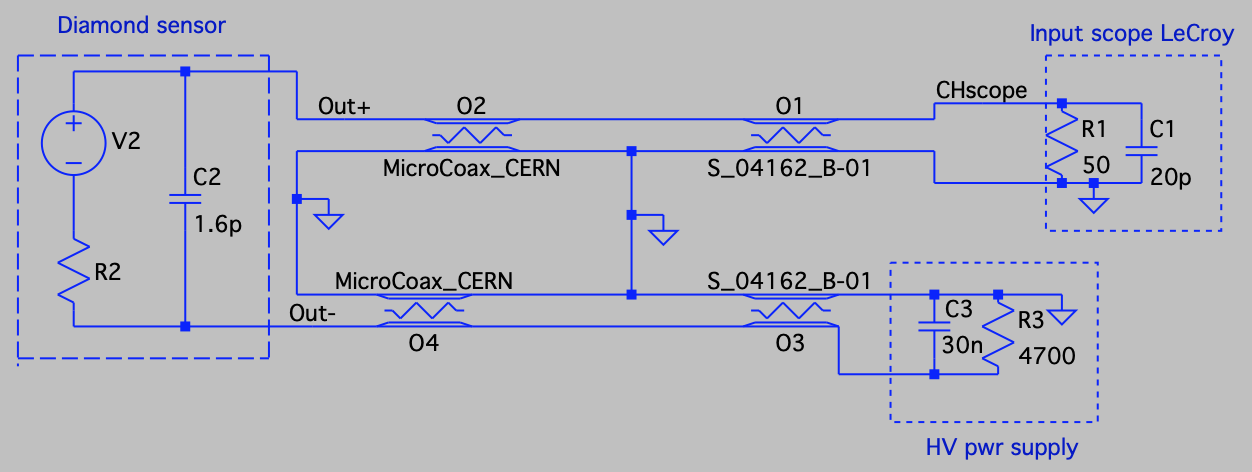}
		\caption{Equivalent circuit for diamond detector irradiated by ultra-short and intense high-energy electron pulses. The diamond detector is represented via the components in the dashed box on the left. Note that this diagram is constructed to investigate the voltage pulse observed on the oscilloscope, the DC bias of power supply is omitted.}
		\label{fig:diagram}
	\end{center}
\end{figure*}

\subsection{Diagram of the circuit}
\label{subsec:diagram}

To be operated as particle detectors, semiconductor devices receive a bias voltage from a power supply. Depending on the circumstance, the system can be represented either as a current source or as a voltage source in the equivalent circuit. As a rule of thumb, if the charging of the power supply is rapid and sufficient, the semiconductor device functions as a current source. On the other hand, if compared with the signal from the device, the charging of the power supply is relatively slow and insufficient, the system can be considered as closed. The energy consumed on the drift of charge carriers inside the semiconductor bulk will induce voltage drop on the electrodes. Consequently, the semiconductor device ought to be regarded as a voltage source.

As shown in Section~\ref{subsec:output}, the maximum of the signal current (voltage divided by 50-ohm impedance of the oscilloscope) is about one ampere as order of magnitude. Conversely, the maximal current provided by the power supply is merely one milliampere~\cite{ref_powersuply}. As a result, the assumption of regarding the biased diamond detector as a voltage source in the case under consideration stands firm particularly for the time window of the initial hundreds of nanoseconds. Therefore, the region of interest in the simulation is defined to be the time window ranging from 0~ns to 300~ns. The diagram of the system is shown in Figure~\ref{fig:diagram}, in which the diamond detector is represented via the components in the left dashed box. The capacitance of diamond detector is, to a good approximation:

\begin{equation}
C = \frac{\epsilon_{0} \cdot \epsilon_{r} \cdot S}{d} = 1.6~\textrm{pF},
\label{eq:capacitance}
\end{equation}
where $d$ and $S$ are the thickness and the cross-sectional area of the diamond crystal, $\epsilon_{0}$ is the vacuum permittivity, and $\epsilon_{r}=5.7$ is the dielectric constant of sCVD diamond. Lossy Transmission Line Model is adopted for coaxial cables. The oscilloscope is modeled using a resistor in parallel with a capacitor. During experiments a 50~$\Omega$ termination is attached in parallel to the channel of oscilloscope whose impedance is set to 1~M$\Omega$. This configuration prevents high current damaging the input circuit of the oscilloscope. The high-voltage power supply is modeled based on its data sheet~\cite{Bepo}. 

\subsection{Resistance of diamond}
\label{subsec:resistance}

\begin{figure}
	\begin{center}
		\includegraphics[width=7.5cm]{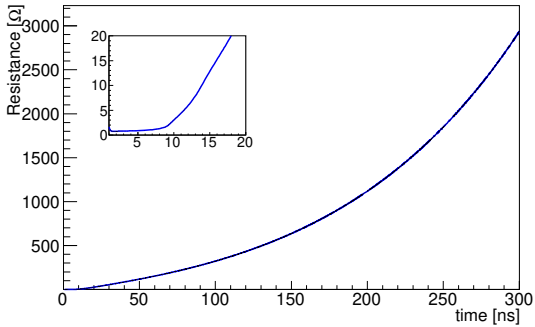}
		\caption{Resistance of the diamond detector obtained from TCAD simulation for the case of V$_{\text{bias}}$ = 100~V. The inset shows the value starts out around 1~$\Omega$.}
		\label{fig:res}
	\end{center}
\end{figure}

Using the simulated evolution of the voltage on electrode V$_{\text{load}}$(t) and the current I(t) in TCAD-Sentaurus, the resistance of the diamond is obtained as R(t) $=$ [V$_{\text{bias}}$ - V$_{\text{load}}$(t)]~/~I(t). One example of the resistance evolution is shown in Figure~\ref{fig:res}, in which the value of resistance initially is about 1~$\Omega$ then starts to increase after about 10 \textmu s. This trend can be accounted for as discussed for the voltage evolution that charge carriers initially drift until the screening effect comes into play. A cross-check on the initial value of the resistance (when all charge carriers are drifting) is done as follows. For diamond treated as an intrinsic semiconductor, the resistance is approximately given by

\begin{equation}
R = \frac{d}{S} \cdot \frac{1}{q_{e} \cdot (n_{n} \mu_{n} + n_{p} \mu_{p}) }  = \frac{d^2}{q_{e} \cdot (N_{eh} \mu_{n} + N_{eh} \mu_{p}) },
\label{eq:resistance}
\end{equation}
where $n_{n}~(n_{p})$ is the concentration of the negative (positive) charge carriers, $N_{eh}$ the total number of e-h pairs, and $\mu_{n}~(\mu_{p})$ the mobility of the negative (positive) charge carriers. The area $S$ cancels out. For an order of magnitude estimate, We assume that e-h pairs are confined with uniform concentration in a cylindrical volume corresponding to the ionization by the incoming beam. Using the value of mobility in Section~\ref{subsec:property}, Equation~\ref{eq:resistance} leads to $R \approx 1~\Omega$ for the resistance of diamond right after irradiation, in agreement with the simulated one (Figure~\ref{fig:res}).

\section{Validation of the simulation approach}
\label{sec:alpha}

\begin{figure*}
	\begin{center}
		\includegraphics[width=15cm]{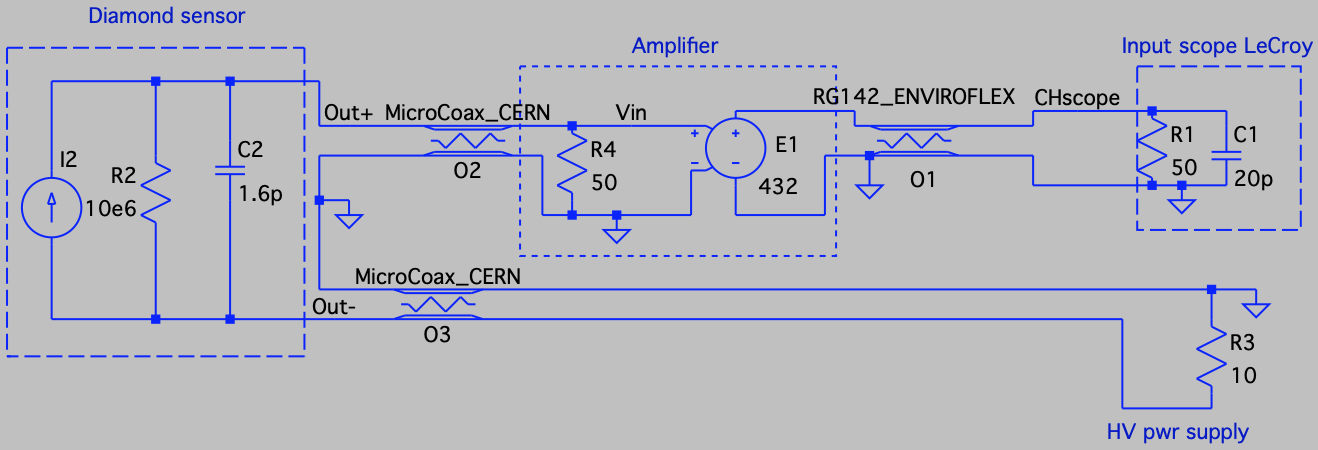}
		\caption{Equivalent circuit for TCT measurement of diamond's transport properties in LTspice. The diamond detector is represented via the components in the left box; the current source includes the active part of the HV power supply.}
		\label{fig:diagram2}
	\end{center}
\end{figure*}

\begin{figure}
	\centering
	    \includegraphics[width=7.5cm]{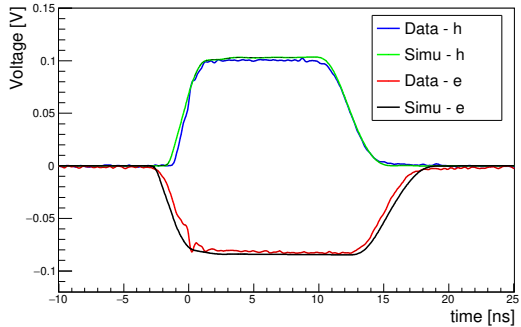}
	    \caption{Voltage observed by the oscilloscope as a function of time. Upper half: hole-induced current; lower half: electron-induced current. The experimental data (blue and red curves) of the TCT measurement are from Ref.~\cite{ref_cali}. The green and black curves are results of the two-step simulation.}
	\label{fig:TCT}
\end{figure}

The new two-step simulation combination (Sentaurus + LTspice) proposed for the investigation of diamond-based radiation monitors must be validated in a well-understood process. The available result in the aforementioned measurement~\cite{ref_cali} on the transport properties of charge carriers in diamond using the transient-current technique (TCT) is a suitable candidate for the validation of the parameters describing the properties of the diamond detector in TCAD.

In the TCT measurement, incident particles are $\alpha$-particles from a 5-kBq $^{241}$Am source with an energy of 5.486~MeV. A detailed geometry layout of the experimental setup, including the collimator and its distance between radiation source and diamond sensor, is implemented in FLUKA, which leads to a simulation result that 90.6\%~(4.97~MeV) of the energy is deposited in diamond bulk. A simplified piece-wise step function that approximates the simulated distribution of deposit energy is input to TCAD-Sentaurus. The injection is carried out along the center of the cylinder (Figure~\ref{fig:geometry}) where a finer mesh is made. 

Moreover, due to the relatively small amount of charge carriers liberated in TCT measurement, the diamond detector is modeled by a current source as shown in Figure~\ref{fig:diagram2}, with a resistor in parallel set to 1~M$\Omega$ due to the low concentration of carriers. The impedance of the power supply CAEN DT1471ET, providing the $\pm$150~V bias voltage is represented by a 10~$\Omega$ resistor. In the experimental setup, a RF amplifier Particulars Am-02A is inserted between the diamond detector and the oscilloscope LeCroy WAVEPRO960 to amplify the small signal from the diamond detector. The gain of the amplifier is measured to be 53.4~dB for signals of positive polarity (electron-induced current) and 52.7~dB for signals of negative polarity (hole-induced current). Owing to the 3-GHz wide bandwidth of the amplifier (even larger than that of the oscilloscope), it is represented by a \textit{voltage dependent voltage source} in the LTspice code with no loss on high frequency components. Coaxial cables RG142 of length three meters are used between amplifier and oscilloscope.

The simulated results of voltage across the oscilloscope input as a function of time, from the two-step approach are demonstrated in Figure~\ref{fig:TCT}, overlaid with the experimental data taken under bias voltage $\pm$150~V. From the the space-time evolution of the charge carriers in TCAD-Sentaurus, the transportation process can be described as follows. One type of charge carriers get collected by the nearby electrode swiftly while the other type of charge carriers drift through the entire thickness of diamond crystal that determines the width of the signal shown in  Figure~\ref{fig:TCT}. The remarkable agreement between the numerical simulation and the measurement, on both the amplitude and the width of the signal, verifies that the transportation process of charge carriers and the effect of the electronic circuit on signal propagation, revealed by the simulation, interpret the experimental data to a great extent.

\section{Results \& discussion}
\label{sec:results}

\begin{figure}
	\centering
	    \includegraphics[width=6.5cm]{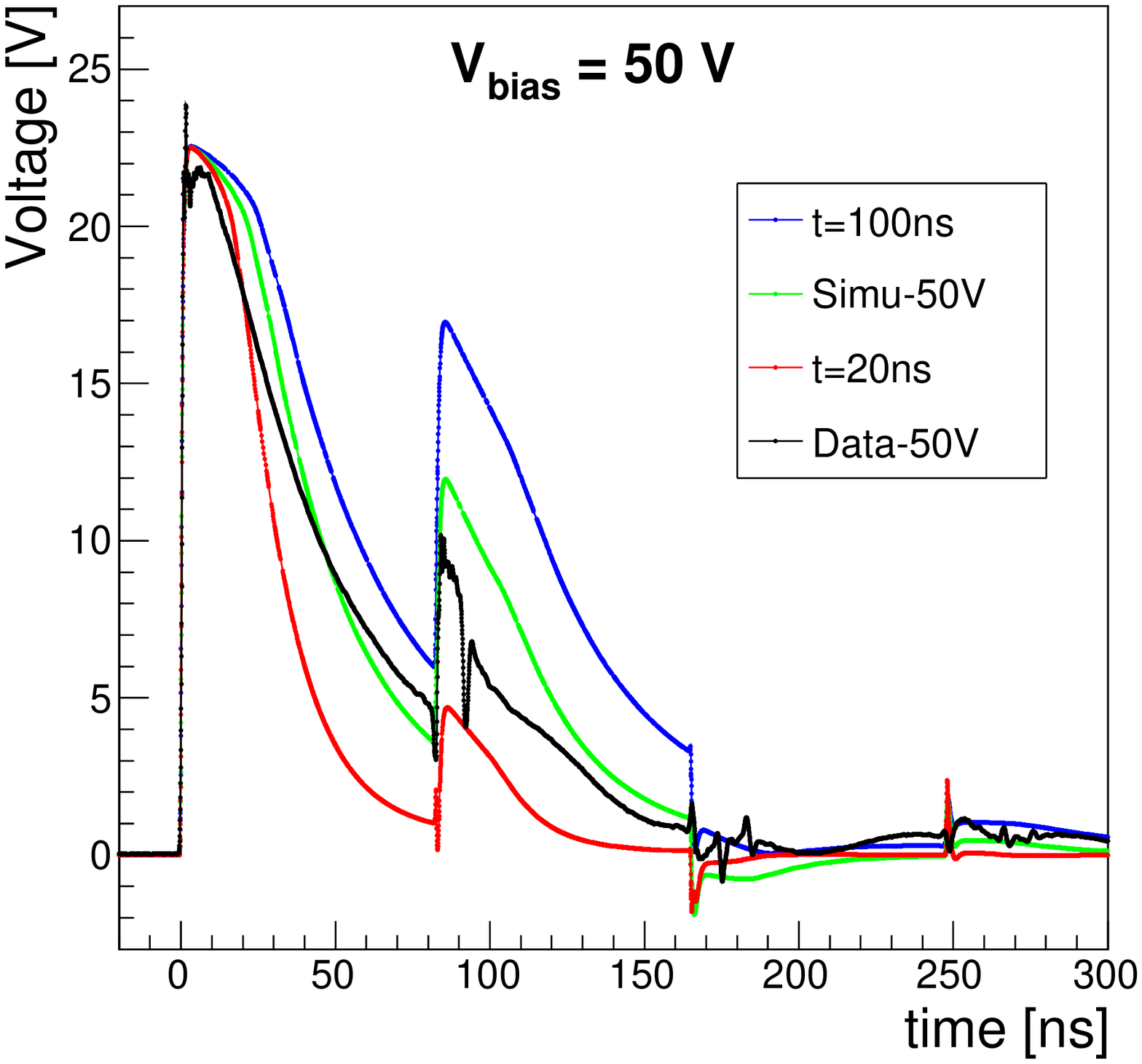}
	    \includegraphics[width=6.5cm]{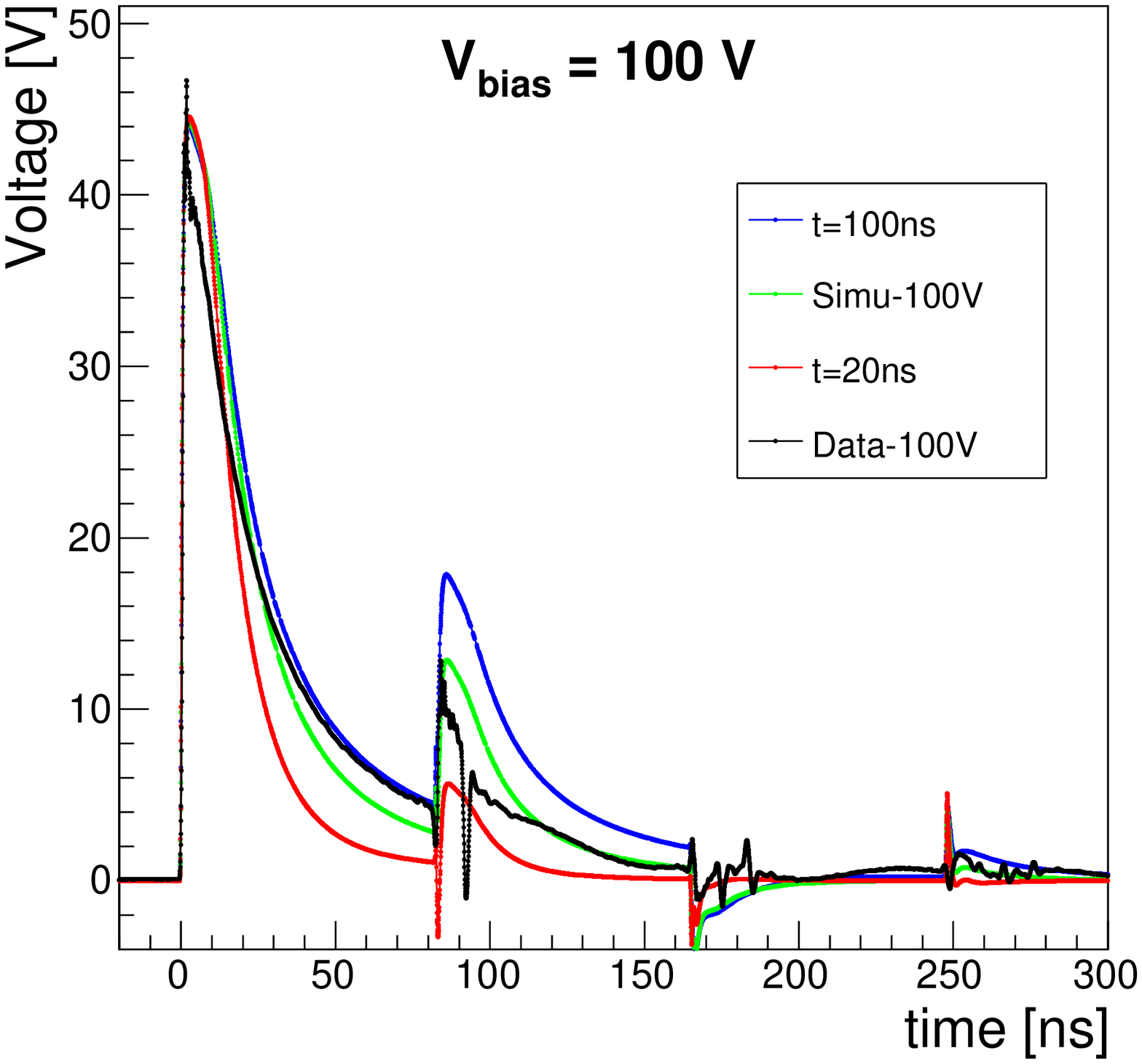}
	    \includegraphics[width=6.5cm]{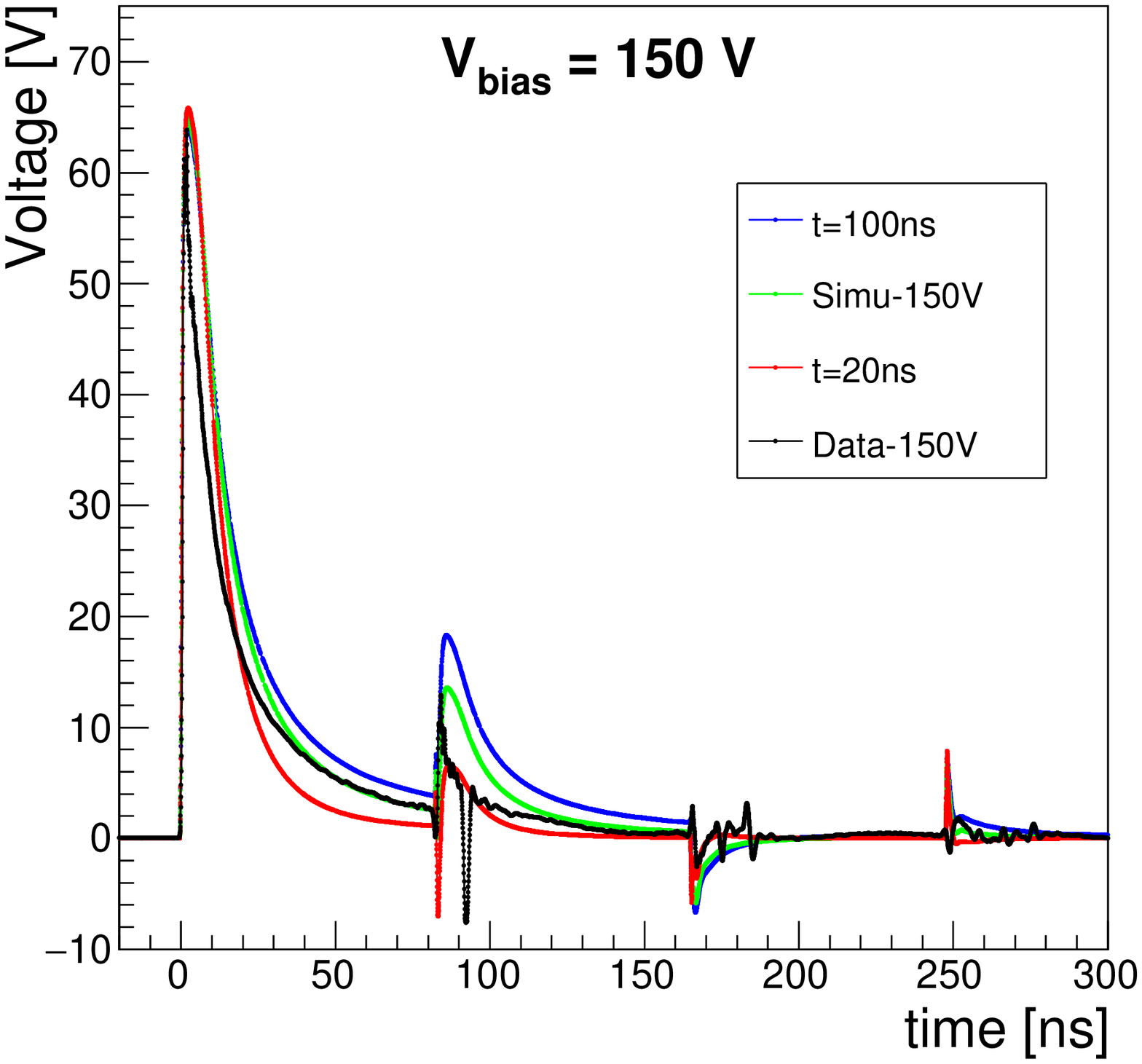}
	\caption{Voltage across the 50~$\Omega$ load, as a function of time, observed by the oscilloscope. Response of the diamond detector to 35~pC electron bunches with 50~V bias voltage (top), 100~V bias voltage (middle), and 150~V bias voltage (bottom). Green curves indicate the results of simulation, black curves indicate the results of measurements. Blue and red curves are results of simulation repeated with different lifetime (t = 100~ns, 20~ns).}
	\label{fig:final}
\end{figure}

After the proof of concept using TCT measurement, results of the simulation on the diamond detector's response to ultra-short and intense high-energy electron pulses are presented in Figure~\ref{fig:final} and compared with preliminary experimental results. In addition, the results of simulation repeated with different lifetimes of charge carriers (20~ns and 100~ns) are also shown in the figure for comparison. 

The amplitude of the signal, as seen by the oscilloscope across the 50 $\Omega$ load, is determined by the impedance of each component in the circuit (diamond sensor, oscilloscope, HV power supply, and cables). In particular, the low resistance of diamond detector right after the irradiation, due to the large concentration of mobile charge carriers, leads to an almost full reception of the signal amplitude (V$_{\text{bias}}/$2) by the oscilloscope.  

In the current experimental layout, the HV power supply has a relatively low impedance for fast signals. The time variable impedance of diamond detector mismatches with the impedance of cables as well. Due to these impedance mismatches, reflections exist in the signals, referring to those oscillating secondary peaks occurring every $\sim$85~ns. This time interval is determined by the length of the cable between diamond detector and power supply. It is worth noting that the reflections could be mitigated using a more symmetrical layout, e.g., increasing the impedance of the HV power supply by inserting a 50~$\Omega$ resistor in series. 

The long tail of the signal is due to the screening effect of space charges as discussed in Section~\ref{subsec:output}, which delays the collection of charge carriers at the electrodes. The slope of the tail is mainly determined by the lifetime of charge carriers. It depends on the dominant recombination mechanism, which may change as a function of concentration, temperature, and time. The minor differences in the peak shape and decay slope indicate the interest of further measurements and test of alternative recombination models in the simulation. 

Assuming collision energy loss for the incident electrons, a 35~pC bunch traversing 0.5~mm thick diamond liberates $5.9 \times 10^{12}$ e-h pairs, in other words, an expected signal charge $9.4 \times 10^{-7}$~C. A measure of the collected charge is obtained by integrating the spectrum and dividing it by the 50~$\Omega$ impedance, leading to a result of $\sim$3$\%$ of the expected signal charge, which indicates that most charge carriers recombine. Charge collection is clearly non-linear in this case. In contrast, in the case of TCT, owing to the absence of screening phenomenon, all charge carriers are collected.

\section{Conclusions}
\label{conclusions}
To decipher the nature of the output signal of diamond detectors under ultra-fast, high intensity radiation bursts, a two-step numerical simulation has been established. Fair agreement between the result of the numerical simulation and that of preliminary experimental data is observed, on both the amplitude and the shape of the response pulse. The observed differences suggest further investigation of the recombination effects and models. The simulation reveals the underlying mechanism of charge collection in diamond bulk and the transmission effect on fast electrical signals. The proposed two-step combination of simulation packages is a flexible and resource-effective approach in investigating both the physics of the devices and the complication related to their external circuits.

\section*{Acknowledgements}
The authors are grateful to the colleagues of the Belle II group at INFN-Trieste and of Elettra Sincrotrone Trieste for stimulating discussions and sharing of preliminary experimental results. We would also like to thank Nicola Zampa for useful suggestions on coding-related issues in TCAD-Sentaurus. 





\bibliography{mybibfile}




\end{document}